\documentclass[aps,pra,twocolumn,floatfix]{revtex4-1}

\pdfoutput=1

\usepackage[utf8]{inputenc}
\usepackage{physics,amssymb,amsmath}
\usepackage{bm}
\usepackage{graphicx}
\usepackage[usenames,dvipsnames]{xcolor}
\usepackage[colorlinks,bookmarks=false,citecolor=blue,linkcolor=cyan,urlcolor=blue]{hyperref}
\usepackage{todonotes}
\usepackage{soul}
\graphicspath{{./figures/},{./prep/}}

\begin{document}

\title{Ground state bistability of cold atoms in a cavity}
\author{B.~G\'abor, D.~Nagy, A.~Dombi,  T.~W.~Clark, F.~I.~B.~Williams, K.~V.~Adwaith,  A.~Vukics, P.~Domokos}

\affiliation{Institute for Solid State Physics and Optics, Wigner Research Centre for Physics,  H-1525 Budapest P.O. Box 49, Hungary}

\begin{abstract}
 We experimentally demonstrate an optical bistability between two hyperfine atomic ground states, using a single mode of an optical resonator in the collective strong coupling regime. Whereas in the familiar case, the bistable region is created through atomic saturation, we report an effect between states of high quantum purity, which is essential for future information storage. The nonlinearity of the transitions arise from cavity-assisted pumping between ground states of cold, trapped atoms and the stability depends on the intensity of two driving lasers. We interpret the phenomenon in terms of the recent paradigm of first-order, driven-dissipative phase transitions, where the transmitted and driving fields are understood as the order and control parameters, respectively. The saturation-induced bistability is recovered for infinite drive in one of the controls. The order of the transition is confirmed experimentally by hysteresis in the order parameter when either of the two control parameters is swept repeatedly across the bistability region and the underlying phase diagram is predicted in line with  semiclassical mean-field theory.
\end{abstract}

\maketitle

\section{Introduction}

Cavity quantum electrodynamics (CQED) is an outstanding platform to study non-linear atom-field dynamics and phase transitions in driven-dissipative systems \cite{ritsch_cold_2013,tanji-suzuki_interaction_2011,mivehvar_cavity_2021}. In its natural setting, a CQED system is driven by external coherent sources, e.g. by laser or microwave radiation, meanwhile the energy is dissipated through a number of channels leading to a steady state resulting from a dynamical equilibrium between driving and loss \cite{walls_inputoutput_2008}. One dissipation channel is the coupling of the cavity field to external, freely propagating, spatially well defined modes, which can be efficiently collected for detection. The outcoupled field then affords an indirect observable of the intracavity steady state \cite{wiseman_quantum_2009}, in the sense of continuous weak quantum measurement.

Cavity QED schemes typically involve few degrees of freedom that are relevant to the atom-light interaction. The field is composed of only a single or a few modes, and the interacting atoms can be represented by a few electronic states. The spatial dimension in which the geometrical size of the atomic cloud can grow macroscopic is irrelevant as only the position relative to the cavity mode antinodes matters. Although the intracavity system size is small, in this sense, the continuous measurement of the outcoupled field is a macroscopic observable and so, it is an order parameter of the system. Transitions between steady states can be affected by changing drive parameters and monitored as a macroscopic change in the recorded signal. Such driven-dissipative phase transitions have been discussed and experimentally studied recently in CQED  \cite{nagy_dicke-model_2010,baumann_dicke_2010,arnold_self-organization_2012,schmidt_dynamical_2014,klinder_dynamical_2015,leonard_supersolid_2017,kollar_supermode-density-wave-polariton_2017,fink_signatures_2018}.

In this paper we present a type of first-order phase transition \cite{vukics_finite_2019,brookes_critical_2021,curtis_critical_2021,reiter_cooperative_2020,megyeri_directional_2018,hannukainen_dissipation_2018,rodriguez_probing_2017,casteels_critical_2017} of an ensemble of cold atoms in an optical Fabry-P\'erot cavity \cite{lambrecht_cold_1995,elsasser_optical_2004,culver_collective_2016,kawasaki_geometrically_2019}. Such a physical realizations of CQED systems have a multitude of applications: the cavity can enable sensitive measurement of the atomic dynamics or state at spectroscopic sensitivity below the standard quantum limit for coherent spin states \cite{chen_cavity-aided_2014,zhang_collective_2012},  real-time monitoring of the spatial distribution \cite{niederriter_cavity_2020} or the atom number in evaporative cooling of atoms \cite{zeiher_tracking_2021}. Superradiance decoherence caused by long-range Rydberg-atom pair interactions, too, has been demonstrated by using cavity-assisted measurements \cite{suarez_superradiance_2022}. Another prospect of strongly coupled atom-cavity  systems is given by optical lattice clocks which are based on lasing on a narrow atomic transition within a resonator \cite{muniz_cavity-qed_2021, vallet_noise-immune_2017, hobson_cavity-enhanced_2019, schaffer_lasing_2020}. Finally, the cavity mode can have a dynamical role such that the hybrid atom-photon excitations introduce new features to non-linear optics. For example, in the case of multiple laser drives, the suppression of polariton excitation by quantum interference \cite{yang_interference_2014} and the proof-of-principle of a multiplexed quantum memory based on spin-waves \cite{cox_spin-wave_2019} have been demonstrated. In these systems, cold atoms can be held in a magneto-optical trap (MOT), or loaded into a cavity-sustained optical dipole trap, or be tightly confined in atom-chip based magnetic traps \cite{kohler_negative-mass_2018}. In our experiment, we use atoms in a large magnetic trap \cite{dombi_self-trapping_2021}, and the cavity mode plays both a diagnostic and dynamical role in the observed phase transition.

In the present study, the phase transition involves two control parameters provided by tunable laser drive powers. One of the lasers is used to probe the transmission of the resonator, whose driven mode is coupled to an atomic transition, whereas the other laser effectively repumps the atoms back into the two-dimensional atomic subspace that is coupled to the cavity. With infinitely strong repumper, the well-known scenario of the atomic-saturation-induced optical bistability in an effectively two-level system is recovered \cite{Martini1993Optical,lambrecht_optical_1995,Sauer2004Cavity,sawant_optical-bistability-enabled_2016}.  Decreasing this control parameter, the system still exhibits bistability with the cavity drive strength control parameter, however, the role of saturation is taken over by populating another hyperfine ground state. Ultimately, the bistability develops into the co-existence of two \emph{phases} in which the internal electronic state of the atoms in the laser-cooled cloud is a pure state, namely, one or the other of the two hyperfine ground states. Recently, in this zero repumper limit, the temporal dynamics of the collapse of an unstable phase has been observed \cite{clark_time-resolved_2021}. In this paper we go on to explore the full phase space spanned by the two control parameters and reveal experimental signatures of the phases and the transition between them.

The paper is structured as follows. In Sec.~\ref{sec:two-way}, we present a model system of competing dynamical optical pumping processes to establish a framework in which to describe our experiments. In Sec.~\ref{sec:PhaseDiagram}, the phase diagram is mapped out by solving the mean-field equations of the model and new features of the bistability domain are pointed out. We make a clear distinction with respect to the well-known case of absorptive optical bistability. In Sec.~\ref{sec:experiment}, the experimental scheme is described and the correspondence to the theoretical model is established. Sec.~\ref{sec:dynamical} is devoted to measurements on the long-time behaviour of the system and to the dynamical signatures of the bistability. Both dynamical oscillations and enhanced fluctuations of the order parameter are demonstrated. In Sec.~\ref{sec:hysteresis} we show that adiabatic ramp cycles of the control parameters lead to hysteresis, which is clear evidence of a first-order phase transition in the system \cite{rodriguez_probing_2017}. Finally, we conclude in Sec.~\ref{sec:conclusion}.

\section{Model of two-way optical pumping of atoms in the cavity}
\label{sec:two-way}

\begin{figure}
\includegraphics[width=\linewidth]{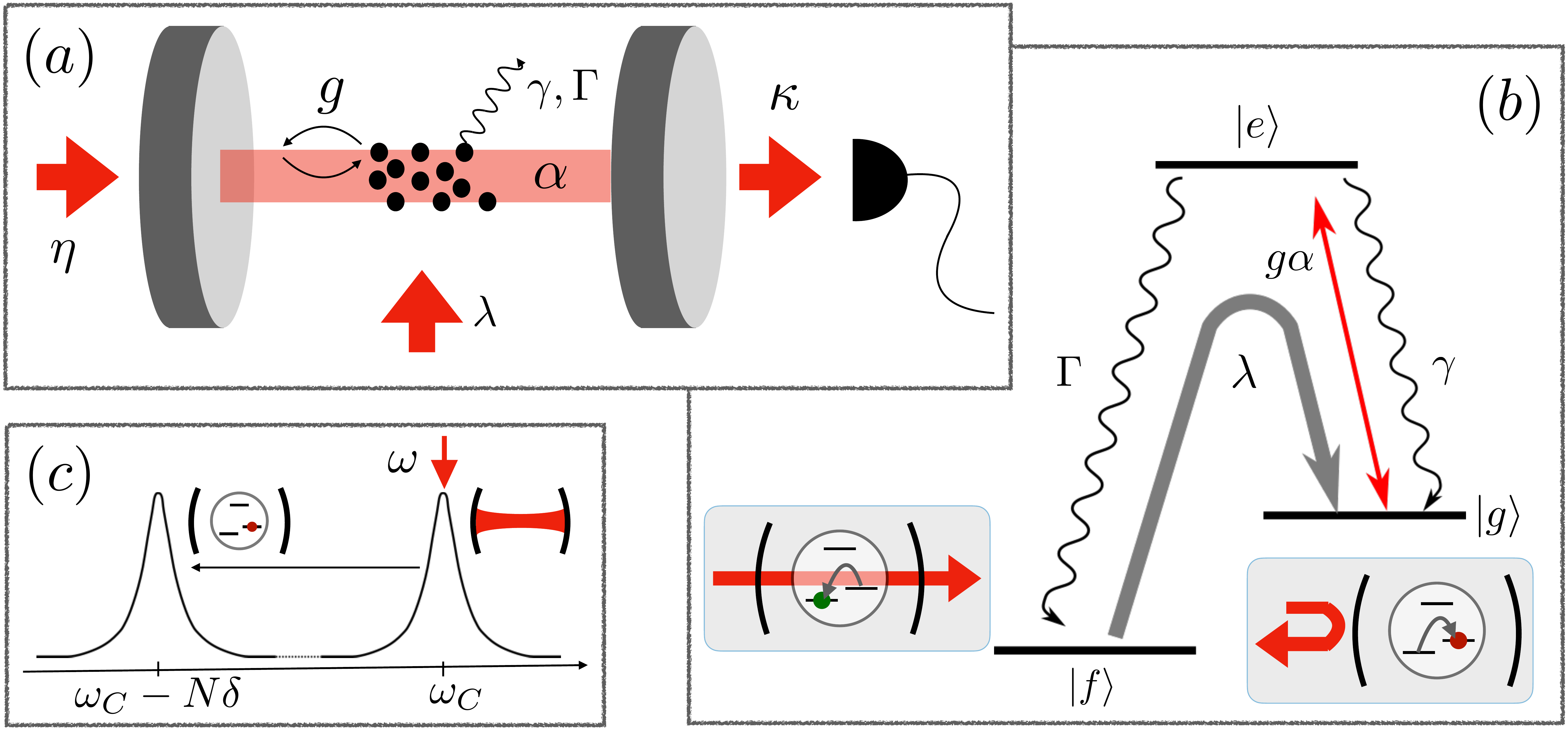}
\caption{(a) The configuration of our CQED scheme. Cold atoms are loaded into a linear cavity and kept in a magnetic quadrupole trap. The cavity is driven with variable effective amplitude, $\eta$, through an incoupling mirror and the transmitted light detected with an avalanche photodiode. The atoms are illuminated from the side by a repump laser of variable power, characterized by the pumping rate, $\lambda$. (b) The relevant part of the atomic level scheme. The transition from the ground state, $|g\rangle$, to the excited state, $|e\rangle$, couples to the cavity mode, resulting in an effective drive amplitude, $g\,\alpha$, where $\alpha$ is the field mode amplitude. The transversely injected repump laser drives the transition from $|f\rangle$ to $|g\rangle$ via other excited states (not indicated). Panels show the cavity transmission accompanying the optical pumping into the states $|f\rangle$ and $|g\rangle$. (c) Atoms in state $|g\rangle$ detune the cavity mode resonance with respect to the laser frequency set on resonance with the empty cavity.
}
\label{fig:scheme}
\end{figure}

We consider ${N}$ atoms interacting with a single mode of a linear optical resonator, as represented schematically in Figure~\ref{fig:scheme}. The cavity mode is driven by a laser with effective amplitude, $\eta$, and angular frequency, $\omega$. This latter is close to the mode resonance, $\omega_C$, such that the detuning $\Delta_C\equiv \omega-\omega_C \lesssim \kappa$, where $\kappa$ is the mode linewidth (HWHM). The cavity field couples to the electric dipole transition $|g\rangle \leftrightarrow | e \rangle$, with coupling constant $g$ (single-photon Rabi frequency). The excited atomic state, $| e \rangle$, decays mostly to $| g \rangle $ with rate $\gamma$ (HWHM), however, a weak decay channel exists to another state, $| f \rangle $, with rate $\Gamma\ll\gamma$. There is a repumper laser illuminating the atoms from the side which performs optical pumping on the atoms back to the state  $| g \rangle $ at a rate $\lambda$. The atomic detuning, $\Delta_A$, is large so the $\Lambda$-atom scheme does not result in electromagnetically induced transparency \cite{joshi_controlling_2003}.

The steady-state of this system manifests a non-trivial phase diagram as a function of the control parameters $\eta$ and $\lambda$. Bistability originates from the competition of the two optical pumping processes, where one of them involves a non-linear cavity-assisted population transfer. A single atom in state $| g \rangle $ detunes the cavity mode resonance by an amount $\delta$. For large-enough atom number, the collective dispersive shift of the atom cloud, $N\,\delta$, can push the mode out of resonance, $|\Delta_C - N\delta| \gg \kappa$, so that the drive $\eta$ is ineffective in exciting the cavity mode. As there is no field in the cavity, the atoms are not excited from the state $| g \rangle $. This solution, dubbed “transmission blockade”, is a steady state. However, it becomes unstable for very large drive strength $\eta$. The Lorentzian cutoff does not eliminate perfectly the transmission. The blockade may break down in a runaway process: for increased cavity drive amplitude, the tiny amount of light infiltrating the cavity excites atoms to $| e\rangle $, which, in turn, results in a reduction of the collective resonance shift and in even more light entering the cavity. This positive feedback amounts to a run-away optical pumping toward the state $|f\rangle$. The extent to which the atoms accumulate in state  $|f\rangle$ depends on the repump rate $\lambda$. For weak $\lambda$, they accumulate; for strong $\lambda$, the atoms are pumped back to $|g\rangle$ and restore the blockading regime. In between, there is a bistability domain where the two steady states can coexist in the form of a statistical mixture. 

The competition between the two optical pumping processes can be described by a semiclassical mean-field model \cite{davidovich_sub-poissonian_1996}. The operator variables in the Heisenberg--Langevin equations are replaced by c-numbers. Let us use the cavity mode amplitude, $\alpha$, and the collective atomic polarization, $M=\sum_i |g_i\rangle\langle e_i |$, where the atoms are indexed by $i=1\ldots {N}$. The populations in the states $|g\rangle$, $|e\rangle$, and $|f\rangle$ are denoted by $N_g$, $N_e$, and $N_f$, respectively. The mean-field equations of motion read
\begin{equation}
\label{eq:meanfield}
\begin{split}
  \dot \alpha &=  (i \Delta_C-\kappa ) \alpha + gM + \eta\,,\\
 \dot M&= (i \Delta_A -\gamma -\Gamma ) M + g\left[N_e-N_g\right]\alpha\,, \\
  \dot N_e&= - g\left[\alpha^* M+ M^* \alpha\right]-
            2 (\gamma + \Gamma)  N_e\,,\\
  \dot N_g&=g\left[\alpha^* M+M^* \alpha \right]+
          2 \gamma N_e  + \lambda N_f \,,\\
  \dot N_f &=  2 \Gamma  N_e - \lambda N_f \,.
\end{split}
\end{equation}
The first equation describes the cavity mode as a driven oscillator which is coupled to the atomic polarization with the strength given by the single photon Rabi frequency, $g$. The second accounts for the atomic polarization due to the cavity field, this process includes saturation nonlinearity. This system of equations is written in a frame rotating with the cavity drive frequency, $\omega$, so the relevant atomic frequency is the detuning $\Delta_A = \omega - \omega_{ge}$, with $\omega_{ge}$ being the transition frequency between states $| g \rangle$ and $| e \rangle$. With these parameters, a single atom induces a cavity resonance shift $\delta=g^2\Delta_A/(\Delta_A^2 + \gamma^2)$, which can be approximated by $g^2/\Delta_A$ in the large atomic detuning regime ($|\Delta_A|\gg \gamma$). In the rest of this paper, we consider the case of resonant driving of the empty cavity, $\Delta_C=0$. Finally, the last three equations represent the evolution of the populations. Besides the cavity-atom interaction and the spontaneous emission term with $\gamma$, here the decay from state $|e\rangle$ to $| f \rangle$ at a rate $\Gamma$ and the repumping of the state $|g\rangle$ from $| f \rangle$ at a rate $\lambda$ are included. This model is a mean-field approach replacing the effect of the individual atoms at different positions and moving with different velocities by collective variables referring only to the internal degrees of freedom.

Although the model is heavily simplified, it is sufficient to capture the main features of the steady-state phase diagram. In particular, the reason why we consider the repumper acting only on the populations of states $\ket g$ and $\ket f$, without creating polarization between them will be explained in Sec. \ref{sec:experiment}.

\section{Steady-state phase diagram}
\label{sec:PhaseDiagram}

\begin{figure}
\includegraphics[width=0.99\linewidth]{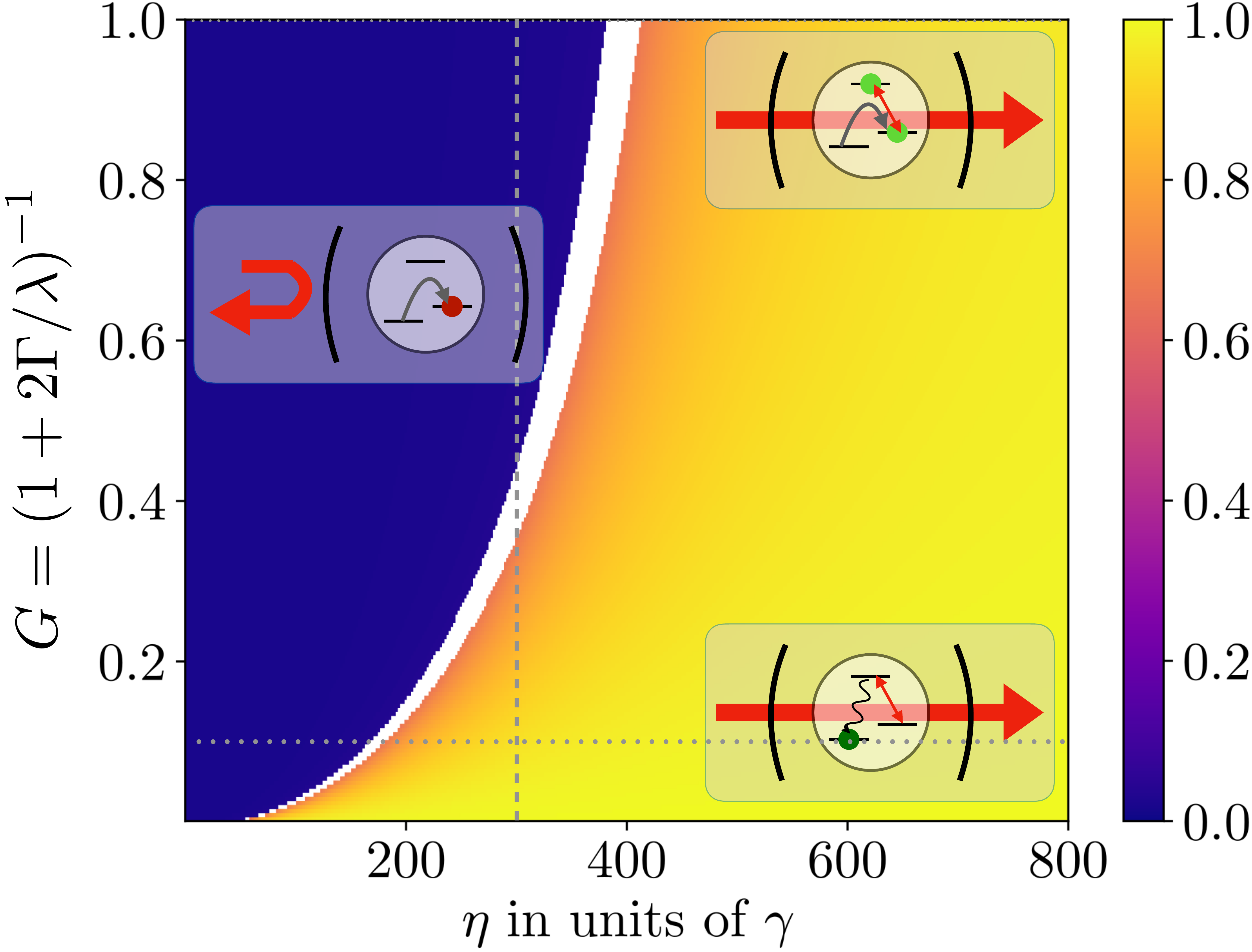}
\caption{Phase diagram of the transmission blockade breakdown in the steady state. The color map shows the cavity transmittance referenced to the resonant empty cavity transmission as a function of the cavity drive amplitude, $\eta$, and repumping rate, $\lambda$. The latter quantity is rescaled with a monotonically increasing function, $G \equiv (1 + 2 \Gamma/\lambda)^{-1}$, which tends to $G=1$ for $\lambda \rightarrow \infty$. The white stripe in the middle corresponds to the domain where the system of equations admits multiple stable solutions. There are distinct phases to the left and right of this boundary which are the blockaded and the bright states of the cavity field, respectively. }
\label{fig:PhaseDiagram}
\end{figure}

\begin{figure*}
\includegraphics[width=\linewidth]{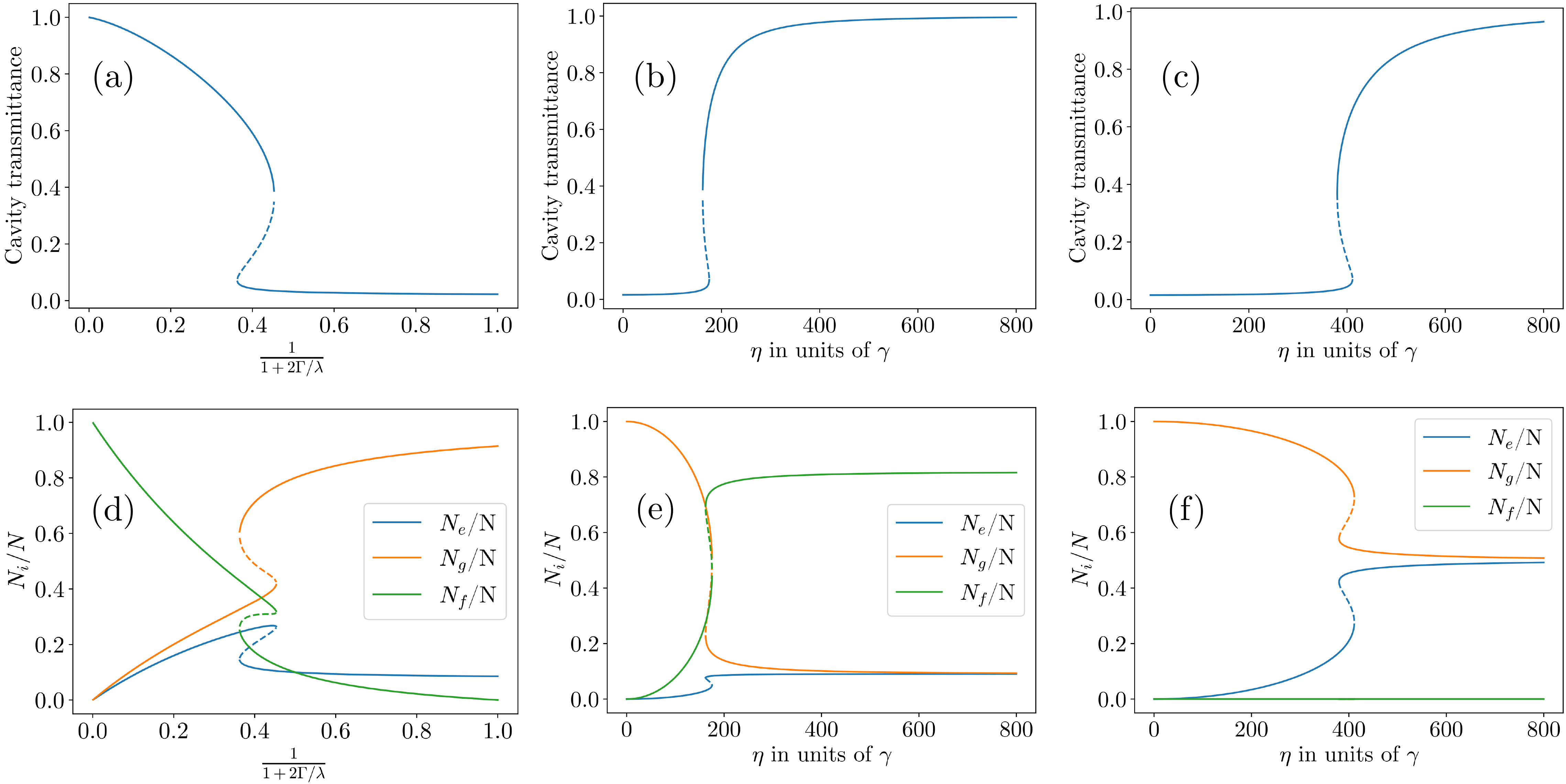}
\caption{
 Cavity transmittance and atomic populations as a function of pumping rates. Transmission is first considered with respect to varying repump rates, with the cavity drive fixed at $\eta=300$, (a). Secondly, we consider transmission as a function of the cavity drive amplitude for fixed repumping rates $G=0.1$, (b), and $G=1$, (c). Similarly, the relative steady-state populations, $N_f$ (green), $N_e$ (blue) and $N_g$ (orange) are plotted with respect to the same pumping rates, (d-f). All the plots show a crossing of the bistability domain, represented by the white stripe, in Fig.~\ref{fig:PhaseDiagram} along a vertical, (a and d), and horizontal (panels (b), (c), (e) and (f)) axis. Where the control parameters, $\eta$ and $\lambda$, give rise to multiple solutions, the solid lines correspond to stable steady states and the dashed ones, unstable solutions.
  }
\label{fig:bistabcuts}
\end{figure*}

The driven-dissipative system defined by Eqs.~(\ref{eq:meanfield}) evolves
towards a steady state that can be calculated by setting the temporal derivatives to zero, and solving the inhomogeneous nonlinear system of equations. Fig.~\ref{fig:PhaseDiagram} presents a color map of the cavity
transmittance in the steady state as a function of the two drive strengths, cavity drive amplitude $\eta$ and repump rate $\lambda$. The transmittance is the transmitted intensity normalized to that of the empty resonator with exactly the same drive $\eta, \omega$. One can clearly observe the blockaded regime for small $\eta$, where the cavity field mode is not populated (dark blue region) as well as a `bright' phase with high transmission (yellow region). These phases are separated by a bistable domain (white stripe), where the system has two stable steady states. These solutions are plotted in Fig.~\ref{fig:bistabcuts} for cross sections of fixed $\eta$ and $\lambda$ values, indicated by dotted and dashed lines in Fig.~\ref{fig:PhaseDiagram}, respectively.

The transmittance exhibits the well-known S-shaped curve as a function of the cavity drive, known from classical optical bistability (cf. Fig.~\ref{fig:bistabcuts}(b) and (c)). There are two stable steady states and one unstable solution. In the present case of a three-level $\Lambda$ scheme, a similar multivalued domain occurs if the repumper power is varied, as shown in panel (a). This highlights the crucial role of the repumper and the third level $|f\rangle$ in the system. The distinctive feature with respect to the well-known case of optical bistability can be revealed by investigating the populations in the three atomic levels   in the steady-state solutions, shown in the bottom row of panels in Figs.~\ref{fig:bistabcuts}(d-f).  In the transmission-blockaded phase, the atoms are dominantly in the state $| g \rangle$, i.e. $N_g \gg N_f,\,N_e$ independently from the repumper and the cavity pump strength. When the blockade is broken down and there is a finite transmittance approaching unity, the populations strongly depend on the repumping rate.

In the low $\lambda$ limit ($\lambda \ll \Gamma$), the cavity photons optically pump the atoms to the other ground state $| f \rangle$, resulting in $N_f \gg N_g,\,N_e$ (see the bottom-right inset scheme in the phase diagram in Fig.~\ref{fig:PhaseDiagram}). This is the interesting limit of bistability, represented by Fig.~\ref{fig:bistabcuts}(e): the two stable steady states correspond to electronic ground states, $| g \rangle$ or $| f \rangle$, with high purity, and the cavity-transmitted photocurrent enables a direct monitoring of which ground state the atoms are in. Such a bistability, dubbed transmission-blockade-breakdown (TBB) \cite{clark_time-resolved_2021}, can be considered the extension of the photon-blockade-breakdown (PBB) phenomenon, known from single- or few-atom CQED \cite{carmichael_breakdown_2015,fink_observation_2017}, to many-atom CQED systems. Whereas with PBB, a large cooperativity ${\cal C}=  g^2/(\gamma \, \kappa)$ is needed on the single-atom level ($g\gg\kappa,\,\gamma)$, for TBB the large cooperativity ${\cal C}=  N g^2/(|\Delta_A| \, \kappa)$ is achieved by increasing the number of atoms (“collective strong coupling regime”).

As a reference, we display the case of classical bistability \cite{Gibbs1976Differential,rosenberger_observation_1983,carmichael_quantum_1986} which is reproduced in the limit of $\lambda\rightarrow\infty$, $G\equiv (1+2\Gamma/\lambda)^{-1} = 1$, when the strong repumper confines the atomic state to the two-level manifold spanned by $| e\rangle$ and $| g \rangle$ (corresponding to the top-right inset scheme in Fig.~\ref{fig:PhaseDiagram})). Fig.~\ref{fig:bistabcuts}(f) shows that the bright cavity phase is connected to a full mixture of the atomic state $N_e \approx N_g$, while $N_f \approx 0$. This means that cavity photons saturate the atoms in the two-level manifold, while state $| f \rangle$ is effectively eliminated from the dynamics by the strong repumper. This model thus reveals that the control parameter $\lambda$ bridges the well-known saturable absorber optical bistability and the much more recent paradigm of first order dissipative phase transitions, that has been shown to be represented by the photon-blockade-breakdown bistability.

In the following, we present experimental results obtained from measurements on a CQED system which is more involved than the above-discussed abstract model. However, we will show that the main features of the interaction are properly captured by the model, and the phase diagram presented in Fig.~\ref{fig:PhaseDiagram} underlies the actual CQED system of the experiment.

\section{Experiment}
\label{sec:experiment}

An ensemble of cold ${}^{87}$Rb atoms was loaded into the mode of a high-finesse resonator by magnetically transporting the atoms from a magneto-optical trap (MOT) to the cavity. After the MOT cycle, the atoms were cooled by polarization gradient cooling down to temperatures of $T \approx 120 \mu$K. Subsequently, they are magnetically polarized by optical pumping into the $(F, m_F)=(2, 2)$ hyperfine ground state to allow capture with a magnetic quadrupole trap. The magnetically trapped atomic cloud was then transported into the cavity by  adiabatically displacing the trap center. The cavity is $l=15$ mm long and the mode waist was $w=127 \mu$m, an order of magnitude smaller than the atomic cloud in this direction. Approximately ${N} \sim 10^5$ atoms were loaded into the mode volume. The cavity linewidth was measured to be $\kappa=  2\pi \times 3.92$ MHz (HWHM), and the single-atom coupling constant was calculated as $g= 2\pi \times 0.33$ MHz on the $(F, m_F)=(2, 2) \leftrightarrow (3,3)$ hyperfine transition of the D2 line. A single mode of the actively stabilized resonator was resonantly driven, $\Delta_C = 0$, with a laser tuned  below the F=2 $\leftrightarrow$ 3 atomic resonance by $\Delta_A=-2\pi \times 29$ MHz. Along with a circularly polarized drive field, $\sigma^+$, the single-atom frequency shift was $\delta \approx 2\pi \times 3$ kHz, such that $N \approx 10^4$ relevant atoms could shift the mode by more than $10 \kappa$ from resonance.

\begin{figure}
\centering
\includegraphics[width=0.66\linewidth]{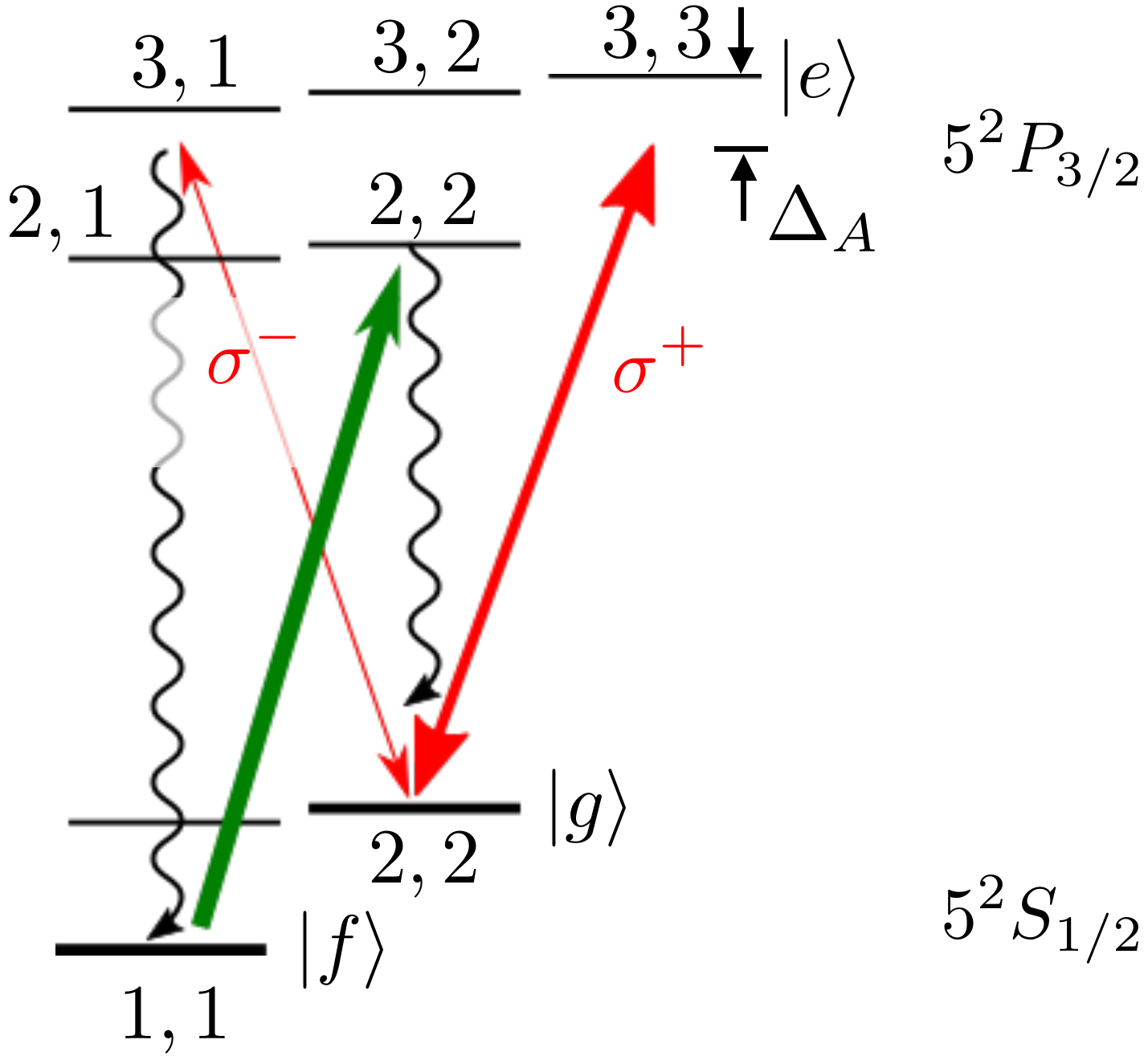}
\caption{The ${}^{87}$Rb levels behind the simplified model in Fig.~\ref{fig:scheme}. Red arrows represent cavity field excitations and the green arrow the repumper. Other states in the hyperfine manifold with smaller magnetic quantum numbers are not shown.}
\label{fig:Rb_levels}
\end{figure}

\begin{figure}
\centering
\includegraphics[width=0.9\linewidth]{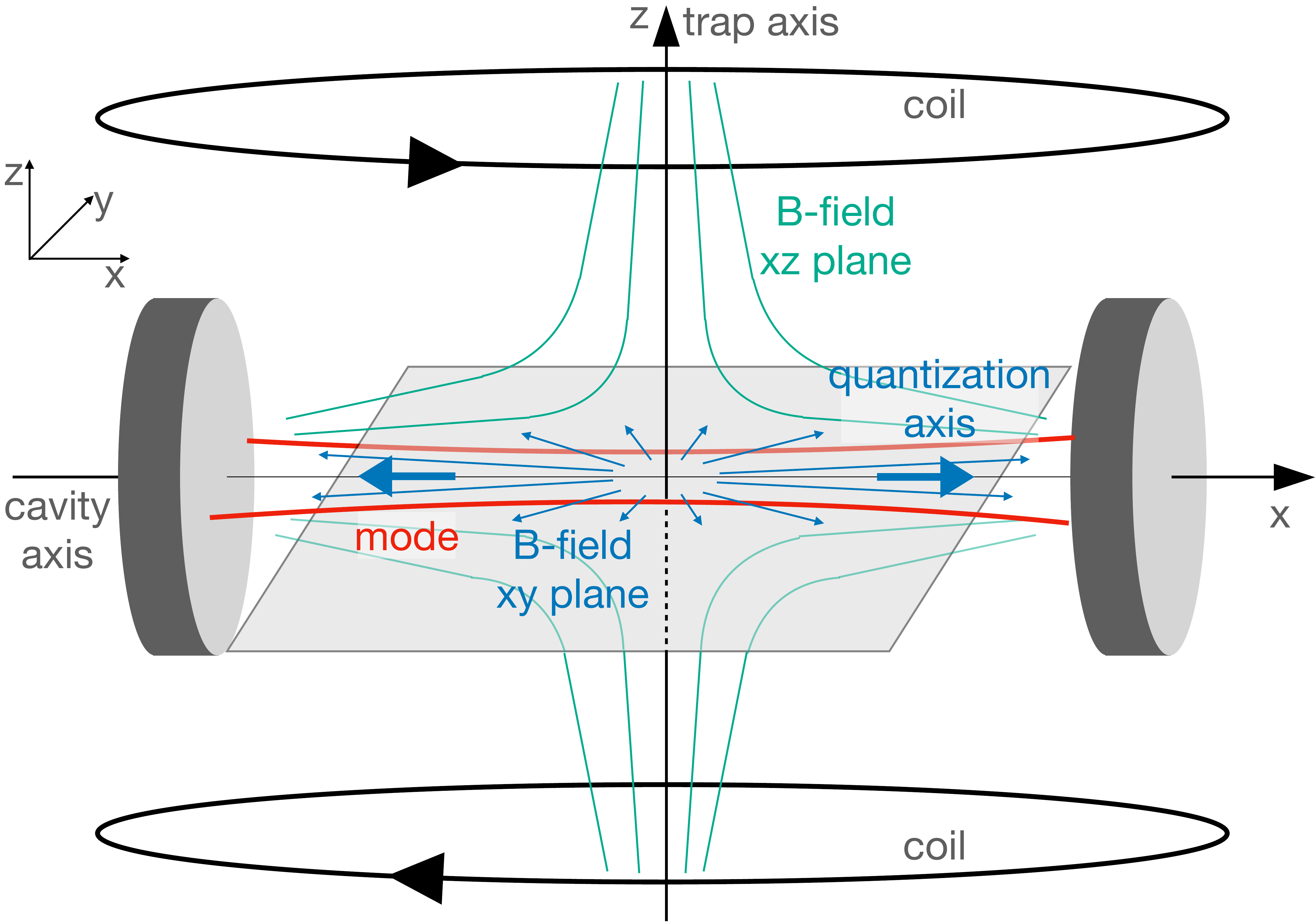}
\caption{Magnetic field lines and orientation with respect to the cavity axis (red lines representing the cavity mode waist in the x-y plane). In the x-z plane, the quadrupole trap creates field lines that bend away from the origin and that are cylindrically symmetric around the z axis (green lines). Perpendicular to this, within the x-y plane of the cavity axis (grey), the magnetic field lines (thin blue arrows) point radially outward. The quantization axis (thick blue arrows), within the cavity mode, is then parallel to the cavity axis but with opposite orientation in the two halves of the mode.}
\label{fig:geometry}
\end{figure}

A mapping between the abstract model of Eq.~(\ref{eq:meanfield}) and the actual level scheme of ${}^{87}$Rb is presented in Fig.~\ref{fig:Rb_levels}, together with the given configuration of laser drives. The state $|g\rangle$ corresponds to the hyperfine ground state $(F, m_F)=(2, 2)$ in ${5}^2{\rm S}_{1/2}$, whereas the excited state $|e\rangle$ is realized by $(F, m_F)=(3, 3)$ in ${5}^2{\rm P}_{3/2}$. This is a closed-cycle transition within the D2 line for $\sigma^+$ circular polarization. As the atoms are in a magnetic quadrupole trap, the magnetic field defining the local quantization axis varies in space. In the plane of the cavity mode, the magnetic field lies in the same plane, pointing radially outward from the trap center which coincides with the center of the cavity (Fig.~\ref{fig:geometry}). Therefore, the magnetic field is oriented, to a good approximation, parallel to the cavity axis within the mode. However, in the two halves of the mode volume, being on the two opposite sides of the mode center, the magnetic field is pointing in opposite directions. Therefore, the circularly polarized cavity drive field is effectively $\sigma^+$ in one half, and $\sigma^-$ in the other half of the mode volume with respect to the local quantization axis.

The $\sigma^-$ polarized field generates transitions from $|g\rangle$ to the  $(F, m_F)=(2, 1)$ in ${5}^2{\rm P}_{3/2}$, which has a much smaller Clebsch-Gordan coefficient than the $\sigma^+$ transition (ratio $\tfrac{1}{15}$). Nevertheless, excitation to the $(F, m_F)=(2, 1)$ implies that the atoms can decay into $(F, m_F)=(1, 1)$ which is the state $|f\rangle$. The decay can also lead to the other hyperfine state $(F, m_F)=(1, 0)$. However, this state can be incorporated into $|f\rangle$. The coupling between the ground-state manifolds $F=1$ and $F=2$ includes a spontaneous emission process in both directions, therefore only the populations, not coherences between the states $|f\rangle$ and $|g\rangle$, $|e\rangle$ play a role. The repumper resonantly drives the transition from the $(F, m_F)=(1, 1)$ ground state to the $(F, m_F)=(2, 2)$ excited state with $\sigma^+$ polarized light, which amounts to an optical pumping into the state $|g\rangle$. This is considered as a population pumping with rate $\lambda$ in equation (\ref{eq:meanfield}).  The other simplification in the semiclassical model is that the population of $|f\rangle$ is loaded from the state $|e\rangle$  rather than introducing additional variables to describe the state  $(F, m_F)=(2, 1)$. The population in this latter is proportional to that of  $|e\rangle$, since both of them are excited by the cavity field from the state $|g\rangle$. Therefore, the crucial dependence on the cavity field intensity and the population in $|g\rangle$ is captured by the model with a phenomenological rate, $\Gamma$, determined previously as $\Gamma= 0.93 \times 10^{-3} \gamma$, by fitting the numerical simulation to the observed transition dynamics \cite{clark_time-resolved_2021}.

The mean-field model, appropriately accounting for the cavity-assisted optical pumping processes,  does not include the atom loss from the trap. The total atom number, ${N}$, in Eqs.~(\ref{eq:meanfield}) is not a conserved quantity. The loss is due to recoil heating, background gas collisions, etc. There are other processes which follow from the dynamics: when the atom is in state $|f\rangle$, the magnetic trap potential vanishes for the $(F, m_F)=(1, 0)$ and is repulsive for the $(1, 1)$ states. Because of the atom loss, the system ultimately evolves into the resonant empty-cavity transmission  on a slow timescale of a few 100 ms (cf. measurement results below). 

\section{Driven-dissipative phase transitions}
\label{sec:dynamical}

The end of the atom transport into the cavity mode defines the time $t=0$. The cavity drive was switched on at $t=3$ ms at the chosen power level. The measurement signal is the cavity transmission recorded by an avalanche photodiode until $t=5$ s. In all the measurements, the atoms were initially in the ground state, $|g\rangle\leftrightarrow(F,m_F)=(2,2)$. Therefore, the transmission at the beginning of the interaction, with $N\sim 10^5$ atoms in state $|g\rangle$, was always suppressed by the dispersive shift of the mode with respect to the fixed drive frequency by more than ten times the linewidth. Depending on the strength of the cavity drive and that of the repumper, this state could be the stable phase or an unstable one, according to the phase diagram in Fig.~\ref{fig:PhaseDiagram}.

\begin{figure}
\includegraphics[width=\linewidth]{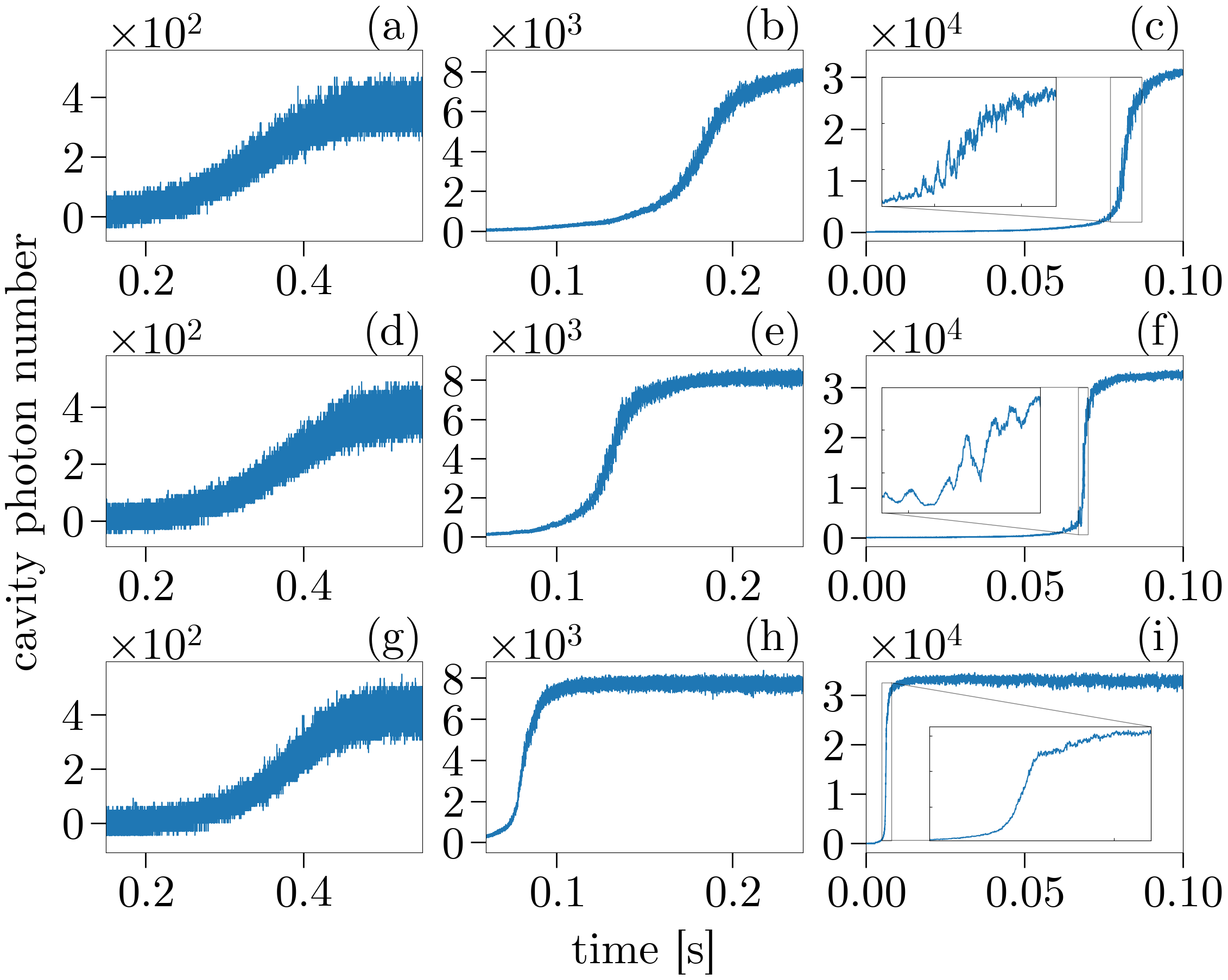}
\caption{The time evolution of the system with respect to varying control parameters, as monitored by the cavity transmission. The cavity drive amplitude increases from left to right ($\eta /\gamma = 25, 117, 236$) and the repumping rate decreases from top to bottom ($\lambda/\gamma = 5.9 \times 10^{-3}, 0.85 \times 10^{-3}, 0.27 \times 10^{-3}$), such that $G=0.76$, $0.31$ and $0.13$, respectively. The dynamics of the transition between the blockaded phase (close to zero transmission) and the transparent steady-state (transmittance close to 1) illustrate different domains of stability in the selected time windows.}
\label{fig:TimeEvolution}
\end{figure}

Figure \ref{fig:TimeEvolution} shows the cavity transmission as a function of time for $3\times3$ different pairs of fixed values of the control parameters $\eta, \lambda$. The cavity drive strength $\eta$ increases from left to right, whereas the rescaled repumping rate $G$ increases from bottom to top.  The left column represents a weak cavity drive compared to the effect of the repumper. According to the phase diagram in Fig.~\ref{fig:PhaseDiagram}, the stable phase is the blockaded one with atoms in $|g\rangle$, which is the initially prepared state of the system. In this case, only atom loss can lead to the transparent cavity state. Indeed, the left column shows that the transition is independent of the repumper strength and takes place on a long timescale of about 350 ms. This timescale can be attributed to the situation that even atoms in state $|g\rangle$ were gradually lost from the trap, due to the above described mechanisms not contained in our idealized theoretical model of Sec. \ref{sec:two-way}. On increasing the cavity drive intensity (middle column of panels), an earlier and faster emptying of the trap can be observed. This indicates that the steady state is still the transmission blockade,  and the cavity drive merely increases the population in the states $|e\rangle$ and $|f\rangle$, thereby speeding up the atom loss. 

Qualitatively different behaviour of the transition is depicted in the right column of panels where the cavity drive is strongest. For the bottom right panel, (i), the repumper drive intensity is so weak that the stable phase is the transparent resonator with atoms in $|f\rangle$. However, the system, initially, is prepared in the other, transmission blockaded phase with all atoms in $|g\rangle$. Before considerable atom loss can take place, the system undergoes a non-linear runaway process to transition into the stable phase. This is clearly the case in (i), and traces of this dynamics can be observed in (h). So the bottom row shows that the transition varies from an atom-loss dominated smooth transition (bottom left panel) to a phase-transition-like switch on increasing the intensity of the cavity drive. This effect has been thoroughly analysed in a recent paper \cite{clark_time-resolved_2021}.

The key new observation is represented mostly by panels Fig.~\ref{fig:TimeEvolution}(f) and (c). Rather than fast, monotonic switching to the stable phase, as in panel (i), stronger repumping leads to an oscillatory transition in (f) and, somewhat less clearly, in (c). The strong dynamical oscillations are indications of the competition of the opposing optical pumping processes. They appear only in a limited range of the control parameters for which, on losing atoms, the system goes into the bistability region of the phase diagram. However, when monitoring the transmitted intensity, the effect of bistability is partly covered by  atom loss. One can unravel the dynamical signatures of the transition which are beyond the effects of the atom loss by analyzing the intensity fluctuations. An alternative method, which we present in the next section, is to vary the system parameters on a timescale shorter than that of the loss.

\begin{figure}
\includegraphics[width=\linewidth]{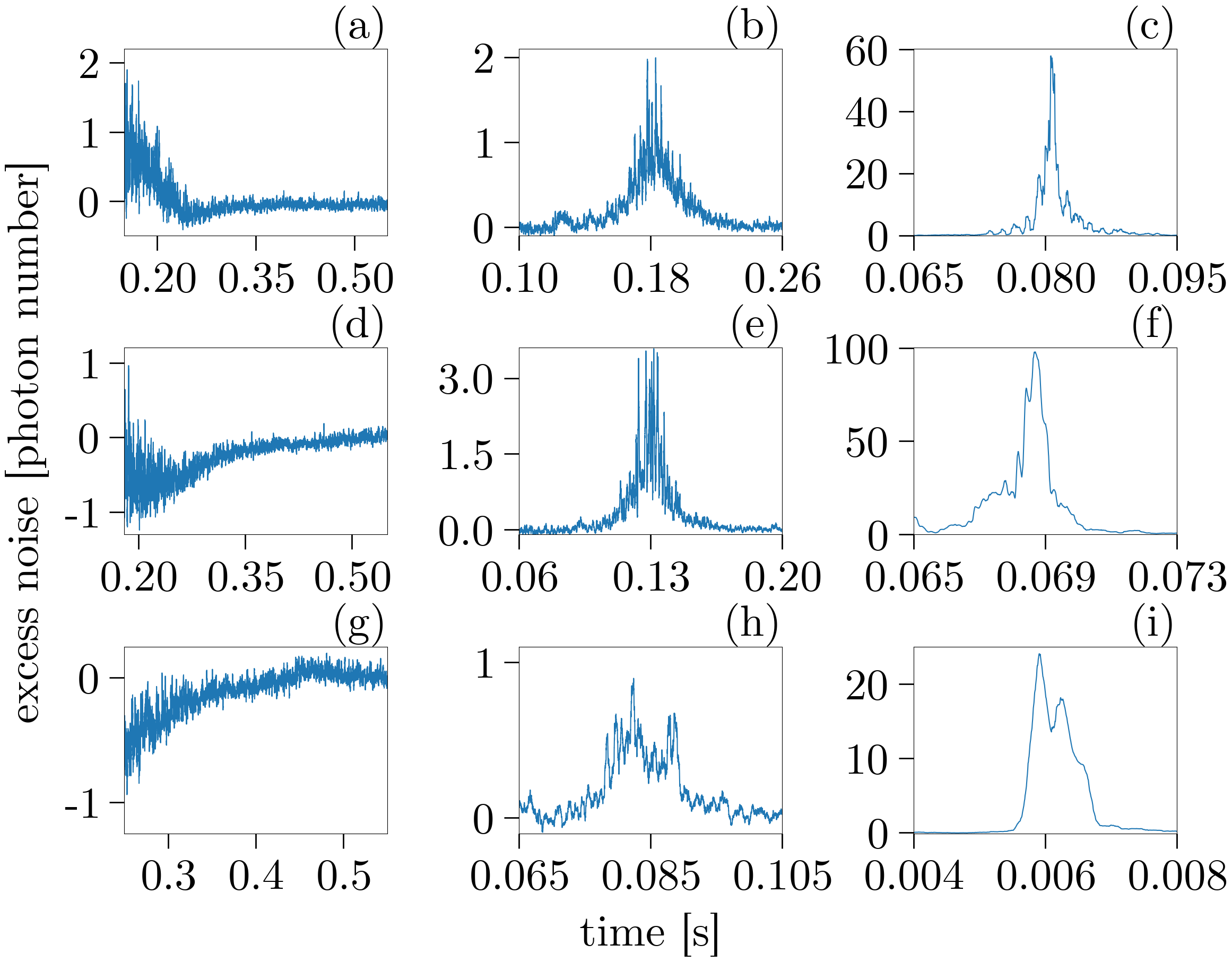}
\caption{The excess noise (in cavity photon number) accompanying the transitions in Fig.~\ref{fig:TimeEvolution}. When the control parameters are in the bistability domain, the single-mode cavity field manifests significantly enhanced fluctuations during the transition between the steady-states.}
\label{fig:ThermalPhotons}
\end{figure}
The bistability is confirmed by the increased intensity of fluctuations in the  detected transmitted signal when the system transitions between  phases. The single-mode cavity field is considered as a displaced thermal (chaotic) state for which the width of the intensity distribution can be characterized by a thermal photon number. This latter can be inferred from the recorded intensity noise following the procedure described in Ref.~\cite{clark_time-resolved_2021}. The excess noise corresponds to fluctuations of the mean-field amplitude, $\alpha$, and is thus beyond the scope of the mean-field model. 

The atom-loss dominated transition to the transparent phase (left column of panels) does not exhibit excess noise during the transition (note that the initial fluctuations including even negative photon numbers in panel Fig.~\ref{fig:ThermalPhotons}(a), (d), and (g) indicate the finite accuracy of the method close to zero mean value of the field, i.e., uncertainty is below 0.5 photon). There is some excess noise generated during the transition with increased $\eta$ (middle column), while significantly enhanced intensity fluctuations accompany the transition for strong cavity drive (right column). An equivalent of 20 and 100 thermal photons characterize the width of the photon number distribution in the cavity mode during the limited time period when the system is in the transmission blockade breakdown and the bistability region, Figs.~\ref{fig:ThermalPhotons}(i) and (f), respectively. This is comparable, but in addition to the Poissonian noise of about 100 photons at the observed mean photon number around $10^4$ (see Fig.~\ref{fig:TimeEvolution}).

\begin{figure}
\centering
\includegraphics[width=0.81\linewidth]{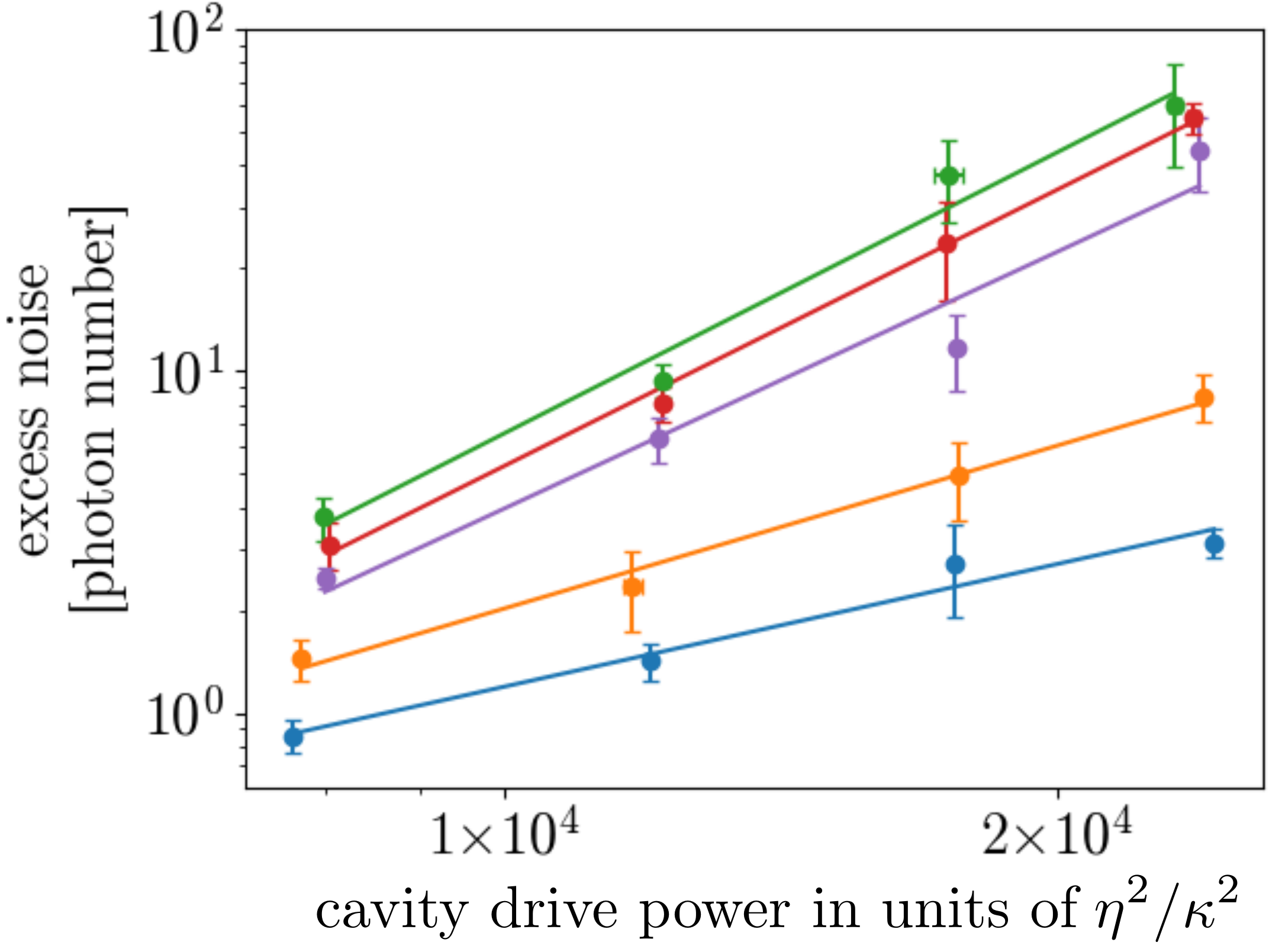}
\caption{The magnitude of the excess noise (in thermal cavity photon number) with respect to the cavity drive power (empty cavity photon number). Different colours correspond to different repumper strengths, $G = 0.127$ (blue), $0.144$ (orange), $0.542$ (green), $0.632$ (red) and $0.708$ (purple).  Each point represents the average of ten measurements. The linear fits in the log-log scale reveal power laws with exponents $1.19\pm0.15$, $1.58\pm0.11$,  $2.73\pm0.27$,  $2.68\pm0.11$ and  $2.49 \pm0.36$, respectively.}
\label{fig:Exponent}
\end{figure}
The enhancement of fluctuations depends systematically on the control parameters. For example, consider the rows of Fig.~\ref{fig:ThermalPhotons} in which only the cavity drive varies. Expressing this enhancement in terms of a thermal photon number, it shows a power law dependence  on the cavity drive as seen on the log-log plot in Fig.~\ref{fig:Exponent} where the drive is also expressed as the  photon number the drive  would generate in an empty cavity. The exponent varies with repumper strength  (numerical values given in the figure caption). A deeper theoretical interpretation of  this experimental observation requires the description of higher-order quantum correlations in the atom-light interaction, which is beyond the scope of the mean-field approximation of the Heisenberg--Langevin equations and will be studied elsewhere.

\section{Hysteresis} 
\label{sec:hysteresis}

As suggested above, the atom number, ${N}$, evolves in time due to loss processes not included in the theoretical model. As such, these changes are not reflected in the phase diagram of Fig.~\ref{fig:PhaseDiagram}. As atoms are being lost, the bistability domain of the diagram shifts toward smaller cavity drive strengths, i.e. the transmission-blockaded phase gradually shrinks. Nevertheless, the atom loss process is slow enough that the multistability of the system is still apparent to fast varying probe light, in the form of hysteresis \cite{rodriguez_probing_2017}.

\begin{figure}
\centering
\includegraphics[width=0.76\linewidth]{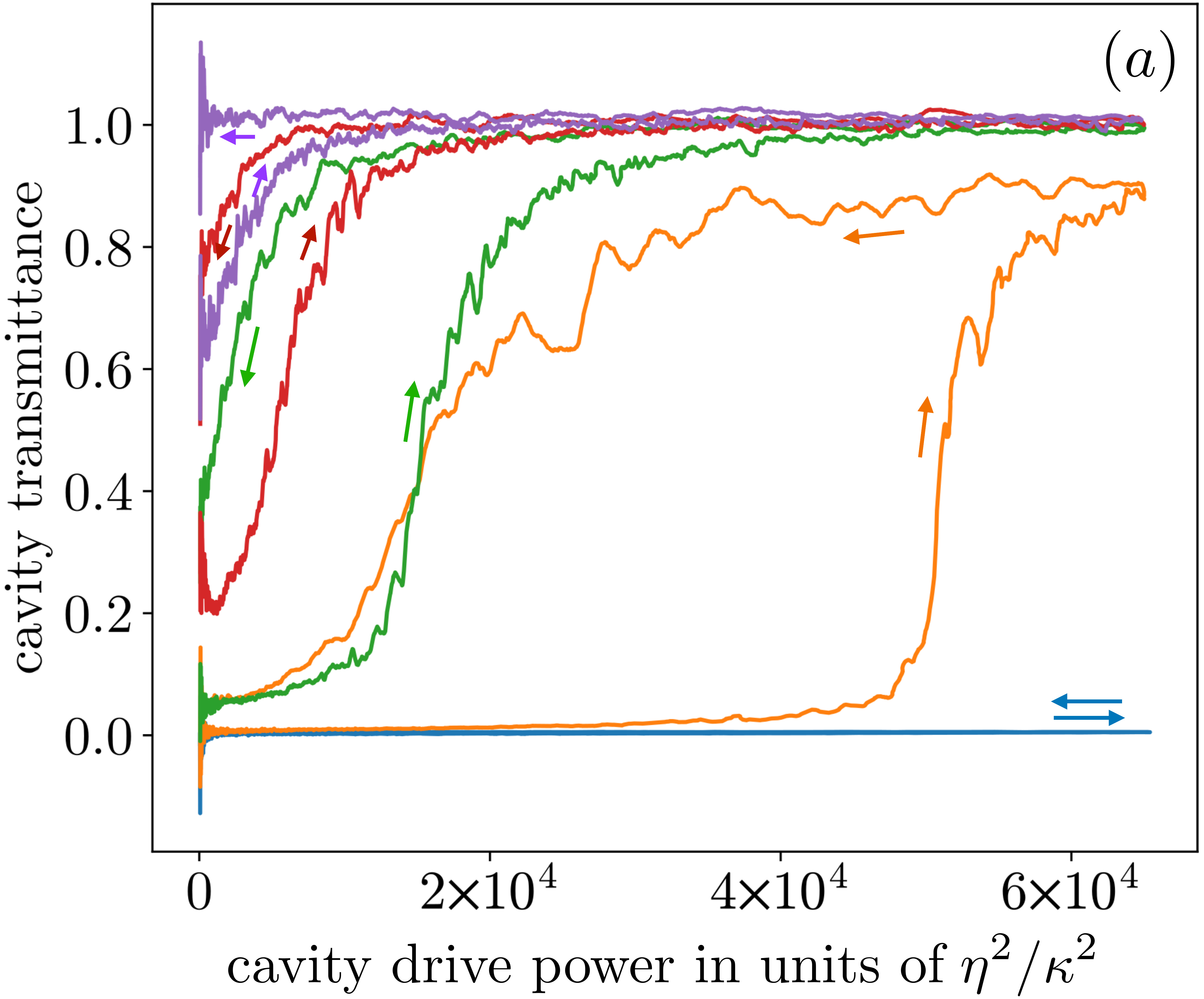} \raisebox{0.5cm}{\includegraphics[width=0.22\linewidth]{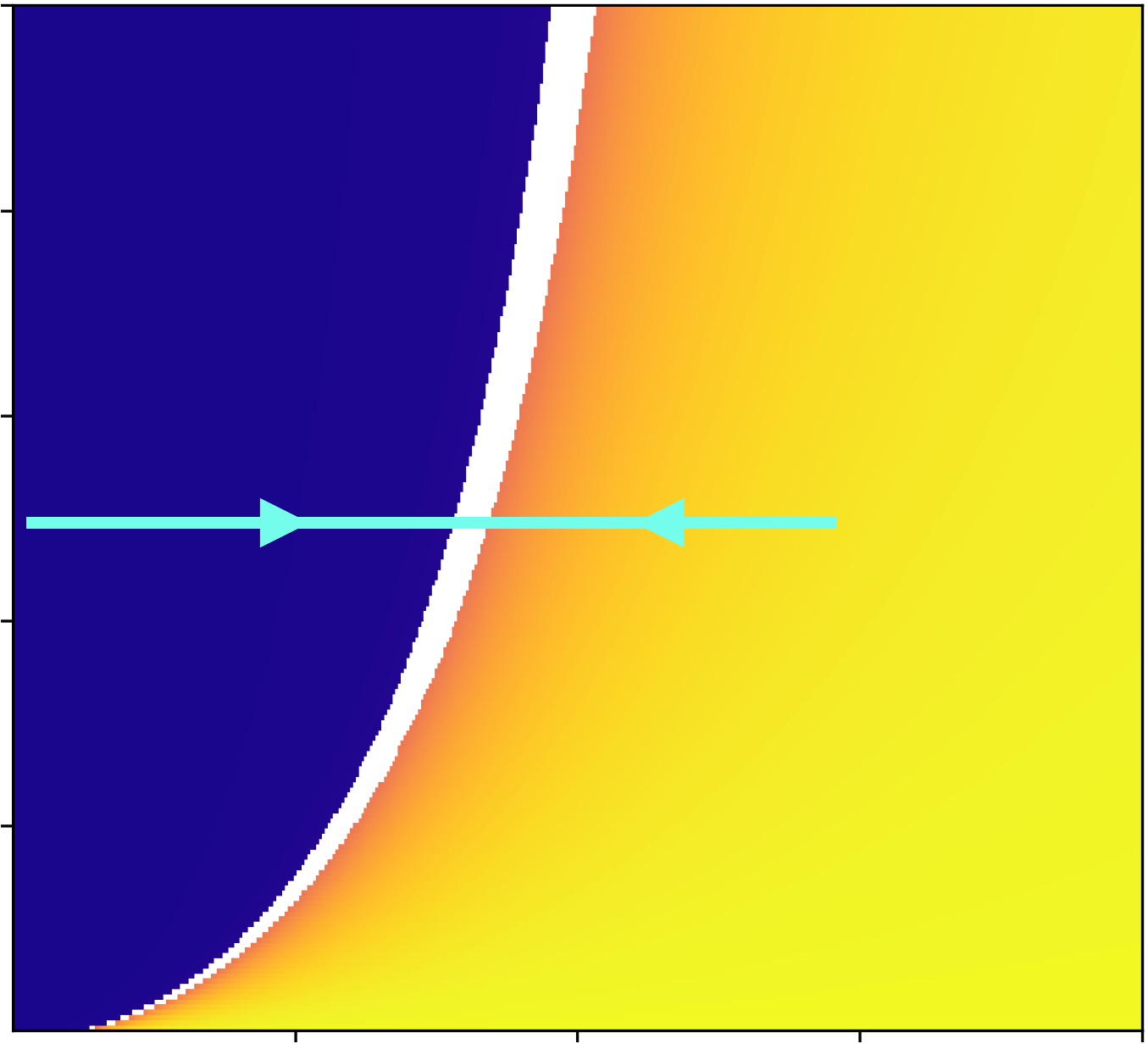}}\\
\includegraphics[width=0.76\linewidth]{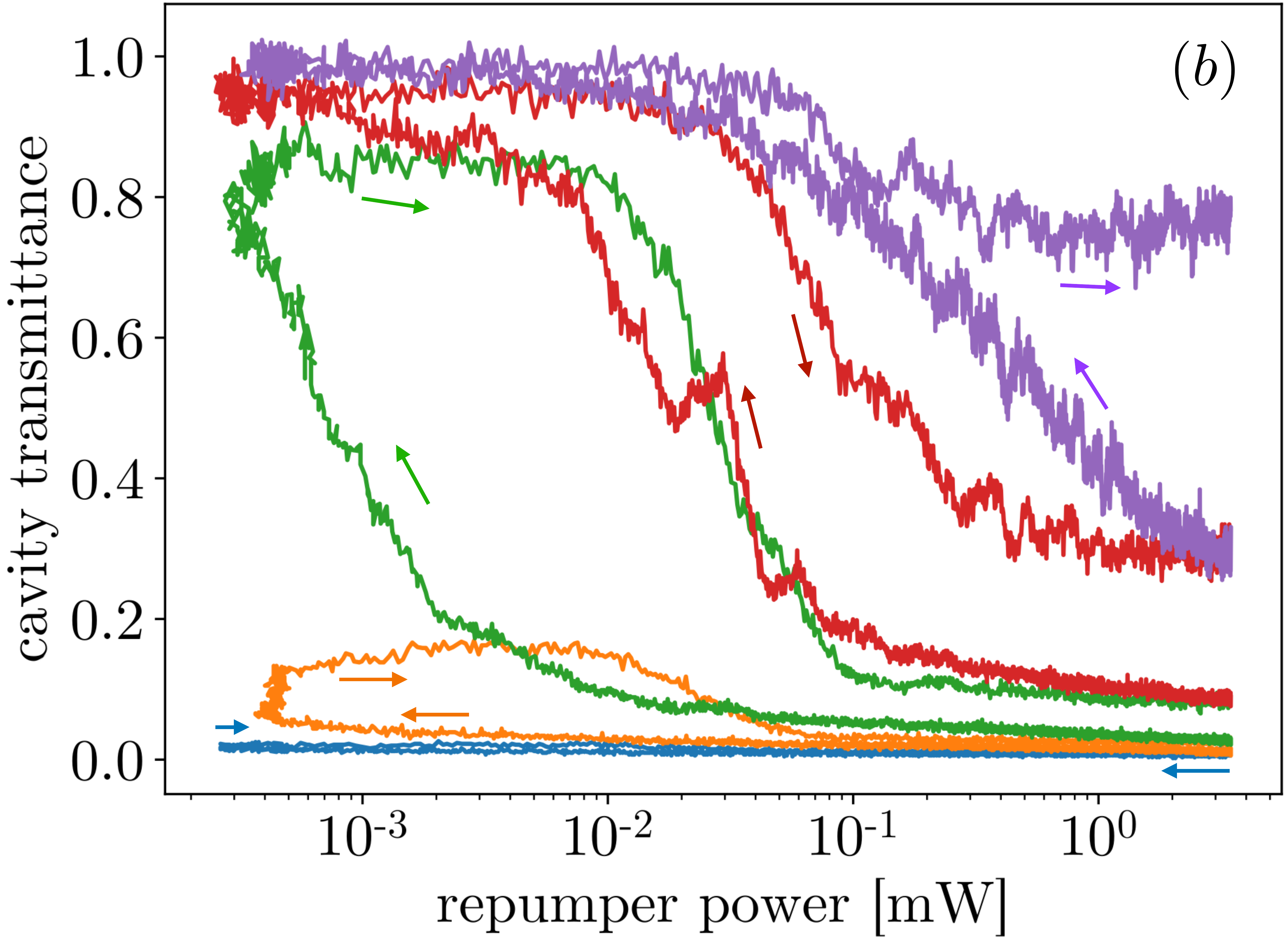} \raisebox{0.5cm}{\includegraphics[width=0.22\linewidth]{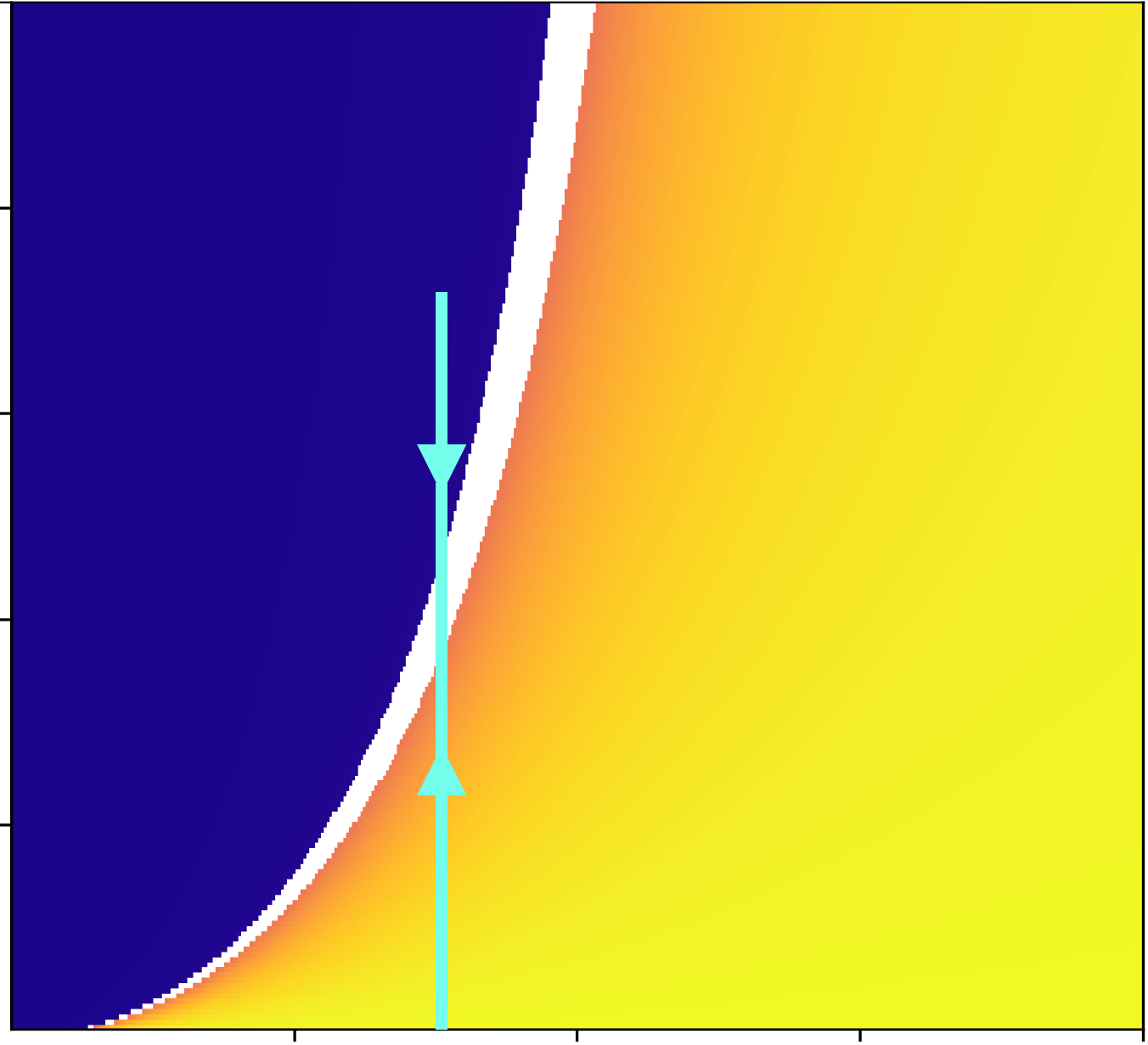}}
\caption{Hysteresis in the cavity transmittance when the cavity drive power, (a), and the repump power, (b), sweep across the bistability domain back and forth five times. The corresponding sweep axis in the phase diagram is shown in the small panels to the right. The temporal order of the ramps is indicated by the blue, orange, green, red and purple lines, respectively and the ramp direction by the arrows. For (b), we use a $\log_{10}$ scale. }
 \label{fig:Hysteresis}
\end{figure}
To this end, the control parameters were swept repeatedly across the bistability domain. The intensities were varied using an acusto-optical modulator (AOM), ramping the drive voltage up and down linearly. For the cavity probe laser, $\eta$, the ramp times were 30 ms up, and 10 ms down, while for the repumper, the corresponding values were 15 and 5 ms, respectively.  Fig.~\ref{fig:Hysteresis} presents the cavity transmission for ramping, (a), the cavity drive and, (b), the repumper intensity five times, while the other control parameter was kept fixed. The atoms were initially prepared in state $|g\rangle$, so the cavity transmission was initially suppressed. The first ramping cycle of the cavity drive did not move the system out of this phase (cf.~Fig.~\ref{fig:Hysteresis}a) because even if this phase becomes unstable for high cavity drive, when the bistability domain is crossed during the ramp, the transition from such a steady state takes place randomly on a long timescale. At this first ramp, it happened to be longer than the sweep period. During the second ramp-up (orange curve), the transition to the transparent phase, atoms in $|f\rangle$, did occur. Hence, during the ramp-down, there is a higher cavity transmission at the same drive strength. This is persuasive evidence of bistability. It is only at the  end of the ramp down period that the repumper transfers the atoms back to state $|g\rangle$, reinstating the blockaded phase. At the beginning of the next ramp up period (green), this is still the ongoing direction of optical pumping until the cavity drive starts to dominate. Accordingly, the corresponding transmission curve (green) is slightly below that of the preceding ramp-down period (orange). This ramp cycle, with the same features, could be observed three more times before the atoms were lost from the cavity. Where we reemphasize that the hysteresis window shrinks for consecutive cycles due to the decreasing number of atoms.

The hysteresis was confirmed by varying the repumper intensity, as a control parameter, with cavity drive fixed. Considering the transition across the bistability domain, the  repumper powers are widely varied that a logarithmic scale is used on the horizontal axis. In Fig.~\ref{fig:Hysteresis}(b) the curve starts at the large repumper limit, where the initial state of atoms in $|g\rangle$ corresponds to the stable phase. Within a sweep cycle, the value of the repumper power was below the bistability domain for short times only. Therefore,  the transition to the states $|f\rangle$ did not happen in the first cycle (blue) but only in the second one (orange). In this cycle, a partial population transfer to the states $|f\rangle$ was accomplished within the ramp-down time and a part of the atoms remained in the blockading state $|g\rangle$. The cavity transmission increased noticeably, but to a value well below the empty cavity reference.  At subsequent cycles, the atom number was smaller due to loss, and a full transfer from the state $|g\rangle$ to $|f\rangle$  has been achieved during the period where the repumper intensity was decreased  below the bistability domain. At these smaller atom numbers, on the other hand, the mode frequency shift did not reach the level necessary to suppress the transmission. Therefore, the red and purple curves do not go down to zero for strong repumper at the right side of the plot. Nevertheless, hysteresis was clearly observed in these cycles, implying the presence of bistability.

\begin{figure}
\centering
\includegraphics[width=0.95\linewidth]{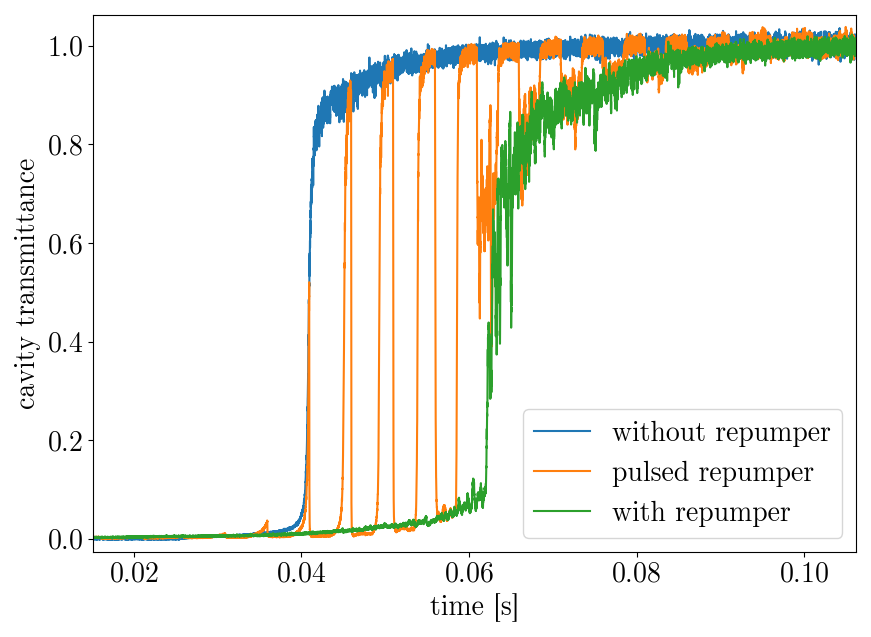}
\caption{The time evolution of the cavity transmission without repumping (blue), with repumping (green) and with pulsed repumping (orange).
When the repumper is switched off, the system starts to evolve in a runaway process toward the bright phase at a given time near $t=0.04$s. With the repumper on however, the system undergoes the transmission blockade breakdown at a later time than without it. As the repumper is strong in this case, the transition to the high transmittance phase, with atoms in state $|f\rangle$, takes place only when the atom number is significantly reduced due to other loss processes. With pulsed repumping, the transition occurs every 5 ms, because the repumper brings the atoms back from state $|f\rangle$ to state $|g\rangle$ (the blockaded phase): switching between the two hyperfine ground states.
}
\label{fig:RepumperRole}
\end{figure}
Finally, we performed a measurement in order to outline the role of the repumper and to detect the atomic state in the transparent phase. The repumper was pulsed between zero and a large value, $G=0.44$, with a period of 5 ms (on/off ratio 1). The time evolution of the cavity transmission is plotted in Fig.~\ref{fig:RepumperRole}, where the blue curve represents the evolution of the system without repumping, while green gives the transmission with constant repumping, for the same value of $G$. These configurations correspond qualitatively to the (i) and (c) panels of Fig.~\ref{fig:TimeEvolution}, respectively. In the prior case, the system with atoms in $|g\rangle$ is prepared in a phase which is unstable at finite cavity drive and without repumper. Therefore the system switches to the stable phase in a runaway process at a random time. With repumper on, the blockaded phase is stabilized to some extent, and the runaway transition is delayed until fewer atoms are present, due to loss. The observed curve for the pulsed repumper demonstrates that the atoms can be transferred back to the state $|g\rangle$ by means of the repumper. The considered timescale is not so long as to have significant atom loss during the period of the plot. Therefore, the blockade can be reinstated repeatedly, following the pulse sequence exactly. This shows that the cavity transmission blockade was broken down by shelving the atoms into the state $|f\rangle$ rather than by an atomic saturation effect.

In addition, this measurement served for the calibration of the model parameter, $\lambda$, characterising the AOM-controlled repumper intensity. When there is a sudden increase of the cavity transmittance (repumper is off), the magnitude of its change gives information on the depletion of the population in $|g\rangle$. One can safely assume that these atoms are accumulated in the state $|f\rangle$. On switching on the repumper, from the initial slope of the transmittance drop, it is possible to deduce the rate of change in the population of $|f\rangle$. It is given by $-\lambda N_f$, according to the last term in Eqs.~(\ref{eq:meanfield}), from which the rate $\lambda$ can be obtained.
 
\section{Conclusion}
\label{sec:conclusion}

We have experimentally demonstrated bistability in a cold atom-cavity QED system, where the steady states correspond, dominantly, to hyperfine ground states. Having explored the runaway pumping processes involved, we described the phenomenon in terms of a driven-dissipative phase transition, with two optical driving intensities as control parameters and cavity transmission as the order parameter of the system.
Crucially, by exploring different combinations of optical pump intensities, we showed that the steady state of the system depends on the history. This observed hysteresis, in both control parameters, not only confirms the bistability but that the transition is a first-order effect. In fact, for high pumping intensity in one of the control beams, we recover the original, widely known, optical bistability, such that our system encompasses this effect as a special case.

Considering future directions, we note that the system size is characterised by the cooperativity, i.e., the collective coupling strength between the atomic cloud and the cavity mode. In our system, the cooperativity, ${\cal C}$, was about 100 which is comparable with the one reached in the circuit QED systems with single artificial atoms \cite{fink_observation_2017}. However, in this $\Lambda$ atom scheme,  the cooperativity can be increased by the number of atoms, so one can better approach the thermodynamic limit. One possible solution is to use Bose condensed gases, for which the steady-states would be entirely quantum in all degrees of freedom. Beyond this, the observed effect is also a promising step towards realising first-order quantum phase transitions. As the cavity transmittance is associated with hyperfine states, which can be coherently manipulated with microwave radiation, the system shows analogy with single-atom based quantum switches \cite{davidovich_quantum_1993} and quantum birefringence systems \cite{turchette_measurement_1995}. Within a many-body systems, the observed effect suggests a pathway for bringing microscopic quantum effects to a mesoscopic system size. 

\section*{Acknowledgements}

We thank J. Fort\'agh, \'A. Kurk\'o and N.~N\'emet for useful discussions. This work was supported by the National Research, Development and Innovation Office of Hungary (NKFIH) within the Quantum Information National Laboratory of Hungary. 

\bibliography{TBB2}

\end{document}